\documentclass[mathleft
]{an}
\usepackage{graphicx}
\usepackage{rotating}
\usepackage{lscape}
\usepackage{longtable}
\usepackage{times}
\begin{document}

\sloppy

% The following seven commands are intended for editorial usage and should be ignored by
% the author(s).
\Pagespan{789}{}% Document's page range.
% If second parameter is left empty, the last page is computed automatically.
\Yearpublication{2011}%
\Yearsubmission{2010}%
\Month{11}%
\Volume{999}%
\Issue{88}%
% \DOI{This.is/not.aDOI}%

\title{Presumable European aurorae in the mid AD 770s were halo displays}

\author{D.L. Neuh\"auser\inst{1}
%Example
%for footnote, note the usage of the \texttt{fnmsep}
%command as separator between institute number and footnote mark}
\and R. Neuh\"auser\inst{2} 
%\thanks{Corresponding author: \email{rne@astro.uni-jena.de}}
}

\titlerunning{Halos in the mid AD 770s}
\authorrunning{Neuh\"auser \& Neuh\"auser}

\institute{
Schillbachstra\ss e 42, 07743 Jena, Germany
\and
Astrophysikalisches Institut und Universit\"ats-Sternwarte, FSU Jena,
Schillerg\"a\ss chen 2-3, 07745 Jena, Germany, rne@astro.uni-jena.de
}

\received{20 May 2015}
\accepted{27 July 2015}
\publonline{ }

\keywords{halo displays -- solar activity -- aurorae -- AD 775}

\abstract{The interpretation of the strong $^{14}$C variation around AD 775 as one (or several) solar super-flare(s)
by, e.g., Usoskin et al. (2013) is based on alleged aurora sightings in the mid AD 770s in Europe: 
A {\em red cross/crucifix} in AD 773/4/6 from the Anglo-Saxon Chronicle, 
{\em inflamed shields} in AD 776 (both listed in the aurora catalogue of Link 1962), 
and {\em riders on white horses} in AD 773 (newly proposed as aurora in Usoskin et al. 2013),
the two latter from the Royal Frankish Annals.
We discuss the reports about these three sightings in detail here. 
We can show that all three were halo displays:
The {\em red cross} or {\em crucifix} is formed by the horizontal arc and 
a vertical pillar of light (either with the Sun
during sunset or with the moon after sunset); 
the {\em inflamed shields} and the {\em riders on white horses} were both two mock suns,
especially the latter narrated in form of a Christian adaptation of the antique dioscuri motive.
While the latter event took place early in AD 774 (dated AD 773 in Usoskin et al. 2013), 
the two other sightings have to be dated AD 776, 
i.e. anyway too late for being in connection with a $^{14}$C rise that started before AD 775.
We also sketch the ideological background of those sightings and
there were many similar reports throughout that time.
In addition, we present a small drawing of
a lunar halo display with horizontal arc and vertical pillar forming a cross
for shortly later, namely AD 806 June 4, the night of full moon,
also from the Anglo-Saxon Chronicle; we also show historic drawings of
solar and lunar halo crosses from G. Kirch and Helevius and a modern photograph.
}

\maketitle

\section{Introduction}

Miyake et al. (2012) found a strong variation in the $^{14}$C to $^{12}$C isotope ratio
in two Japanese trees in data with 1-yr time resolution, around AD 775.
For the cause of this variation, they excluded supernovae due to the lack of
any such historic observations (nor supernova remnants) and solar flares due to
the observed $^{14}$C to $^{14}$Be ratio for that event.
The increase around AD 775 is one of three strong fast rises between 1000 BC and AD 1900 
in (Intcal) data with 5-yr time resolution (Neuh\"auser \& Neuh\"auser 2015b),
one of them began in the AD 1790s at the start of the short Dalton minimum,
i.e. due to a decrease in solar activity and wind.

Gibbons \& Werner (2012), Melott \& Thomas (2012), Usoskin et al. (2013), and Zhou et al. (2014)
suggested that the $^{14}$C variation around AD 775 was caused by strong solar activity,
one (or several) solar super-flare(s).
The hypothesis is supported significantly by alleged aurorae 
which should indicate strong activity in the mid AD 770s. 
One of those reports is from China on a presumable aurora 
({\em ten bands of white qi}) misdated to the end of Dec 775,
which was more likely some halo effect {\em above the} 
(almost full) {\em moon} -- and correctly to be dated
to mid Jan 776 (Chapman et al. 2015).

We will discuss here the remaining three European reports from the mid AD 770s in detail,
also considering their ideological background.
This paper is based on previous work (Neuh\"auser \& Neuh\"auser 2014)\footnote{which is
a proceedings paper in German, we give here a free English translation in particular 
of section 23.2 (we also expand the discussion somewhat here) and parts of section 23.3}.
A critical review of historical aurorae from AD 731 to 825 is
given in Neuh\"auser \& Neuh\"auser (2015a).

In connection to the $^{14}$C variation around AD 775,
Allen (2012) suggested that an entry in the Anglo-Saxon Chronicle (ASC),
presumable for AD 774,
could be the key to understand the radiocarbon variation: 
\begin{quotation}
This year also appeared in the heavens a red crucifix, after sunset; 
the Mercians and the men of Kent fought at Otford; 
and wonderful serpents were seen in the land of the 
South-Saxons.\footnote{cited after Allen (2012) from
avalon.law.yale.edu/medieval/ang08.asp, a collation without date correction} 
\end{quotation}

Allen (2012) -- and, following him, also Firestone (2014) -- interpreted 
the {\em red crucifix} as absorbed supernova
(with presumably invisible remnant) and as cause for the $^{14}$C variation.

The conjecture by Allen (2012) cannot be confirmed for several reasons:
A supernova hardly ever appears in the form of a cross or crucifix.
However, SN 1006 was reported with the following wordings by Arabic observers:
{\em its edges were [lines like] fingers ... had four strands bare of extremities}
in the  Mauritanian report (Goldstein 1965, Stephenson \& Green 2003);
{\em it was not circular, but nearer to an oblong.
At its ends, there were lines like fingers.
It showed a great turbulence} (al-Yam\={a}n{\={\i}} from Yemen, Rada \& Neuh\"auser 2015),
{\em it was not round, but rather was elongated;
at its edges were [lines like] fingers.
It showed a great turbulence} (Ibn al-Dayba$^{c}$ from Yemen, Rada \& Neuh\"auser 2015).
These apparent elongations are due to very strong scintillation of a
very bright object ($-7.5$ mag). Given the large brightness,
SN 1006 was then visible for many months on the whole hemisphere, 
while the {\em red cross} is reported only for one evening in one single source.
Furthermore, the remnant would be detectable now by X-rays, $\gamma$-rays,
and/or radio observations, which have lower extinction then the optical
(see e.g. Hambaryan \& Neuh\"auser 2013, who excluded also absorbed supernovae as cause).

Gibbons \& Werner (2012) presented another celestial event,
for AD 776, from the Annales Laurissenses: 
\begin{quotation}
two shields burning with red colour and moving 
above the church itself.
\end{quotation}
and also from the Chronicon of Sigebert of Gembloux:
\begin{quotation}
when the Saxons besieged the castle of Heresburch [Eresburg], 
the glory of God appeared to all, 
surely as two shields burning with the colour of blood 
and making certain motions through the air, as if at war.
\end{quotation}

These two quotations probably refer to the same celestial event,
even though the events reported in Chronicon Sigeberti are connected
there to the place called Eresburg,
while the Annales Laurissenses maiores -- named after the monastery
of Lorch in Germany, from where the oldest manuscript of the 
Annales Regni Francorum (Royal Frankish Annals, RFA) originated --
mentioned the castle Syburg instead of Eresburg, both in today's north-western Germany.
This confusion probably goes back to Chronicon Regino (written about AD 906). 
Gibbons \& Werner (2012) do not discuss this matter, but also
see the same event in both quotations.
Gibbons \& Werner (2012) also note that the colour of red often is
used as portent for battle-related events and that the {\em red crucifix}
is mentioned in connection with the {\em Battle at Otford}, which is
usually dated to AD 776 (Stenton 1970).
They then conclude that there may have been
an {\em extended period of auroral activity}. 
From the fact that those two {\em red shields} were even sighted during
the day, they conclude that the phenomenon was very bright --
{\em if indeed it was a cosmic event} (Gibbons \& Werner 2012).
Hence, they implicitly connect the $^{14}$C variation, as measured
in tree rings, with strong solar activity.

The conclusion by Gibbons \& Werner (2012) is questionable,
because for those two reports ({\em red cross} and {\em two shields})
-- as compared to other reports from that time --
it is dubious whether they refer to aurorae, and they both have to be dated for AD 776,
so that they cannot indicate strong solar activity before AD 775.
There are aurora reports, e.g., in AD 762 from China and in AD 772 and 773 from
the Near East, which are stronger and more trustworthy 
(e.g. Yau et al. 1995, Xu et al. 2000, Neuh\"auser \& Neuh\"auser 2015a).
Furthermore, if those reports would indicate strong solar activity
and, hence, strong solar wind, 
the cosmic ray flux entering the solar system, and then also
the radiocarbon production, should {\em de}crease.
In such a case, then, an extra-solar origin would be more
likely for the $^{14}$C {\em in}crease.

Usoskin et al. (2013) suggest that so much $^{14}$C would have been produced
by solar protons during solar storms that it would be detectable in tree rings since AD 775;
however, $^{14}$C from solar storms was never confirmed in historic tree rings, 
where all the radiocarbon is averaged over one whole year,
e.g. not even during the so-called Carrington event,
a $^{14}$C variation was detected (e.g. Neuh\"auser \& Hambaryan 2014).

By saying {\em the phenomenon seems to have been observed during the day, 
suggesting that it was very bright if indeed it was a cosmic event},
Gibbons \& Werner (2012) implicitly suggest the latter, 
even though that the reports presented by them for AD 776 would
indicate an {\em extended period of auroral activity}
(Gibbons \& Werner 2012), i.e. not a single super-strong solar flare.

Zhou et al. (2014) claim to have found the {\em world-wide super-auroras of the last 11400 yr},
namely an alleged Chinese aurora ({\em white qi above the} [almost full] {\em moon} 
misdated by them for AD 775 Dec (true: AD 776 Jan, Chapman et al. 2015);
Zhou et al. (2014) connect the latter (possibly a halo display) with two European sightings, 
namely the {\em red cross} and {\em celestial lights} in Belgium,
also strongly misdated by them, to construct {\em world-wide super-auroras}.
This claim is falsified in Neuh\"auser \& Neuh\"auser (2015a).
Stephenson (2015) interpreted both the Chinese {\em white qi} and the
European {\em red cross} as possible aurorae, but see below. 

We present the three presumable European aurora sightings in the mid AD 770s 
with context and transmission history in turn in the next
three sections (Sect. 2-4); then we discuss the medieval form 
of narrating mock-suns as a Christian adaptation of the antique {\em dioscuri motive} (Sect. 5); 
we finish with a summary in Sect. 6.

\section{''Red cross/crucifix''}

Usoskin et al. (2013) follow explicitly the {\em superflare hypothesis}
(without citing Gibbons \& Werner 2012):
{\em The AD 775 cosmic event revisited: the sun is to blame},
discussing also some previously suggested historic reports. 
They date the {\em red cross} from the ASC to AD 773, 774, or 776: \\
{\em Allen (2012) interprets the red cross (...)
as an exotic nearby supernova with an unobservable remnant, 
but we interpret this as an aurora} (Usoskin et al. 2013).

Let us discuss the historic source:
The Anglo-Saxon Chronicle (ASC) exists in six main versions (different copies),
all in medieval English (A-F), one of them (bi-lingual F) also in medieval Latin.
The events discussed here are reported in very similar wordings,
dated to AD 774 (version A gives AD 773).
However, AD 776 is the correct date, because events from
AD 754 to 845 are shifted by two years (3 years in A) due to mistakes by copying scribes
(Garmonsway 1953).
Later, events from northern England were added to versions D and E,
so that those are not shifted (Garmonsway 1953).
The reported {\em Battle at Otford}, a small town in southern England,
therefore happened in AD 776 (Stenton 1970, Gibbons \& Werner 2012).

The quotations in the different versions of the ASC read as follows\footnote{manuscripts 
A to E cited in medieval English from asc.jebbo.co.uk/b/b-L.html}:
\begin{itemize}
\item Manuscript A for AD 773: \\
{\em Her obiewde read Cristesmel on hefenum \ae fter sunnan setlgonge;
$\&$ by geare gefuhton Mierce $\&$ Cantware \ae t Ottanforda; 
$\&$ wunderleca nedran w\ae ron gesewene on Subseaxna londe}, \\
which is in medieval English, it includes {\em a red cross/crucifix in the sky/heaven after sunset}, 
the battle at {\em Otford}, and {\em adders} in south Saxon land;
\item Manuscript B for AD 774: \\
{\em Her ooeowde read Cristes m\ae l on heofonum \ae fter sunnansetlgange,
$\&$ by geare gefuhtun Myrce $\&$ Cantware \ae t Ottanforda, 
$\&$ wundorlice n\ae ddran w\ae ron gesawene on Subseaxna lande},  \\
similar to A;
\item Manuscript C for AD 774: \\
{\em Her ooywde read Cristes m\ae l on heofenum \ae fter sunnan setlgange,
$\&$ by geare gefuhtun Myrce $\&$ Cantware \ae t Ottanforda, 
$\&$ wundorlice n\ae dran waron gesewene on Suosexena lande}, \\
similar to A;
\item Manuscript D for AD 774: \\
{\em Her Norohymbra fordrifon heora cyning Alchred of Eoforwic on Eastertid $\&$ genamon \ae belred Molles
sunu him to hlaforde, se ricsade .iiii. winter. $\&$ men gesegon read Cristes mel on heofenum \ae fter sunnan setlgange.
On by geare gefuhton Myrce $\&$ Cantware \ae t Ortanforda, $\&$ wunderlice n\ae dran w\ae ron gesewene on Suoseaxna lande}, \\
similar to A plus northern events;
the report indicates that the sighting was observed by {\em men}
(this term meant male or female person(s) in medieval English, Clark Hall 1960),
i.e. that it was actively seen by some people;
the collations in Garmonsway (1953) and Whitelock (1979) 
show that {\em se ricsade .iiii. winter} means that {\em he reigned for four years}; 
since the text(s) in D (and E) bring(s) first all northern events and then
all southern events, we may conclude that the sighting of the {\em red cross} 
possibly was near the start of their (southern) year, since it is mentioned first;   
\item Manuscript E for AD 774: \\
{\em Her Norohymbra fordrifon heora cining Alhred of Eoferwic on Eastertid $\&$ genamon \ae delred Molles
sunu heom to hlaforde, $\&$ se rixade .iiii. gear; 
$\&$ men gesegon read Cristes mel on heofenum \ae fter sunnan setlangange.
On by geare gefuhton Myrce $\&$ Cantwara \ae t Ottanforda, 
$\&$ wundorlice n\ae dran w\ae ron geseogene on Suoseaxna lande}, \\
i.e. very similar to manuscript D, but {\em gear} instead of {\em winter} (for {\em year}).
\item Manuscript F (bi-lingual in Latin and medieval English) for AD 774: \\
Medieval English (from Thorpe 1861): {\em Her \ae delred Molles sunu rixiun agann on Northymbran. 
\& menn gesegan read Cristes m\ae l on heounu \ae fter sunnan setlegange.
On than ylean geare fuhton Myree \& Centwar\ae~at Ottefordan. 
\& wunderlice n\ae dra w\ae ron gesawene on Suthsexan.} \\
Medieval Latin (from Magoun 1947): {\em Aedelred filius Moll cepit hic regnare,
\& visum est crucis signum in c\ae lo post solis occubitum.
\& eodem anno pugnaverunt inter se Merci \& populus Cantiae apud Otteford,
\& in Sudsexa visae sunt serpentes mirabiles}. \\
These texts say that Aethelred, son of Moll, started to reign over Northumbria (report from the north,
afterwards reports from the south),
that the sign of the cross was seen in the sky after sunset,
and that wonderous neddran were seen in southern Sussex. \\
The Latin translation follows mainly the medieval English in F,
which is similar to D and E, but a bit shorter:
The main difference is that the {\em read Cristes m\ae l}
was reported as {\em crucis signum}, i.e. without the colour of red.
See Fig. 1.
\end{itemize}

Given the 2-yr shift (versions B-F) in ASC dates around that time,
and given that the years (in ASC) 
usually ran from Christmas to Christmas,\footnote{There are also several other
calendar systems used in the ASC. According to the most extreme cases,
the year listed by the authors as AD 776 on their calendar 
could have started as early as our AD 775 Mar 25 (Annunciation Stylus Pisanus) 
or as late as our AD 776 Mar 25 
(Annunciation Stylus Florentinus starting the year on 25 March following our Jan 1).} 
we can date the event to AD 776.
Hence, the {\em red cross/crucifix} event cannot be related
to the (solar or extra-solar) cause of the radiocarbon variation, 
which is seen in tree rings already in AD 775,
so that the relevant radiocarbon was mostly produced before AD 775.

\begin{figure*}
\begin{center}
{\includegraphics[angle=0,height=11cm,width=14cm]{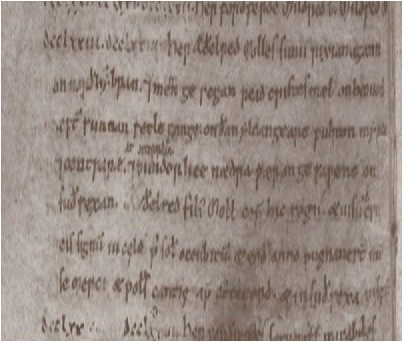}}
\caption{{\bf AD 776.} 
A copy of a small part of the {\em Anglo-Saxon Chronicle} version F 
(bilingual Canterbury Epitome manuscript Cotton Domitian A. VIII folios 29-69
from the British Library, London, UK) in medieval English (top) and Latin (bottom) 
with the entry presumably for AD 774, but in reality with events from AD 776.
The first fully seen line from the top starts with {\em dcclxxiii.dcclxxiiii},
the Latin numbers for the years AD 773 and 774, meaning that there was no report for AD 773,
and that the following text would be for the year AD 774 (correct 776). 
In the 2nd \& 3rd line, we can read {\em \& menn gesegan read Cristes m\ae l on heounu
\ae fter sunnan setlegange}.
At the end of the 4th line from the bottom, 
the Latin translation starts, and one can read 
{\em \& visum est cru-}, which continues at the start of the next
line with {\em cis signum in c\ae lo post solis occubitum}; the p-like sign between {\em c\ae lo}
and {\em solis} stands for {\em post} (e.g. Lutz 1981) meaning {\em after/during} (Niermeyer 1976),
consistent with the meaning of the medieval English \ae fter (Clark Hall 1960).
This is translated to: {\em and it was seen the sign 
of the cross/crucifix in the sky/heaven after/during sunset}
(there is no mentioning of the red colour in the Latin version of F).
This scan is from the facsimile in Dumville \& Keynes (1995);
re-printed here with permission of the publisher Boydell \& Brewer Inc., UK.}
\end{center}
\end{figure*}

The special content of versions D and E (partly also F) of the ASC
usually report about events in northern England:
Would the {\em red cross} seen in southern England (Sussex),
have been truly an aurora, it should also have been seen in
northern England (unless of bad weather).
But there is no such report, not even within
a few years before or after AD 774-776.
There are otherwise three reports about likely true aurorae in the ASC
(from AD 550 to 845), which all come from northern England
(closer to the geomagnetic pole).

From the different versions, the most probable original text
was reconstructed in modern English (e.g. Whitelock 1979, square brackets by us): 
\begin{quotation}
[AD] 776 ([but] 774 [in] C, D, E, F; 773 [in] A)  \\
In this year a red cross appeared in the sky after sunset. 
And that year the Mercians and the people of Kent fought at Otford. 
And marvellous adders were seen in Sussex.
\end{quotation}

For the medieval English {\em Cristesmel/cristes m\ae l/mel} (and the Latin {\em crucis signum}), 
some translations use {\em cross}, others {\em crucifix};
and for the medieval English {\em hefenum/heofonum/heofenum} (Lat.: in c\ae lo or coelum),
some translations use {\em heaven}, others {\em sky}.

For the {\em wonderful serpents} (see Allen 2012) reported for the same year, 
Usoskin et al. (2013) claim, referring to Dall'Olmo (1980): \\
{\em Serpents often feature in descriptions of aurorae, 
reflecting the sinuous movement of auroral structures.} \\
Apart from the fact that a {\em red cross} does not look sinusoidal,
both these elements -- the {\em red cross} and the {\em serpents} --
were already combined to one event in the aurora catalog of Link (1962),
as cited by Usoskin et al. (2013).
Dall'Olmo (1980) listed {\em serpens} ({\em snake}) and
{\em serpens igneus} ({\em fiery snake}) both for
aurorae as well as for sporadic meteors. 
While Dall'Olmo (1980) listed {\em crux} ({\em cross}) also under
aurorae, he also remarked that this {\em medieval term ...
may refer also to paraselenic features}, 
which was not considered in Usoskin et al. (2013).

What was called {\em serpents/snakes} in Link (1962), Dall'Olmo (1980),
Allen (2012), and Usoskin et al. (2013), were actually {\em adders} 
(in medieval English: nedran/n\ae ddran/n\ae dran) (Whitelock 1979), 
which really exist in southern England.\footnote{There is only
one location in all the different ASC versions, 
where serpents or snakes were mentioned, 
namely in an entry for AD 1037 (version E):
{\em They put them [some people] in dungeons wherein 
were adders [nadres] and snakes and toads ...},
which shows that the ASC does make a difference between snakes
(serpents) and adders; the likely true aurorae in the ASC are
never described as snakes nor adders.}

The report about this {\em red cross/crucifix} (even without the so-called {\em serpents/snakes})
was misinterpreted as aurora several times before. 

Jeremiah (1870) 
may have been the first by quoting the {\em red crucifix}
from the ASC and commenting: {\em The auroral hypothesis 
may satisfactorily apply}; he dated it to AD 743, which may possibly
be due to an older ASC edition.\footnote{Jeremiah (1870) 
gives also the year AD 744, 
which goes back to Florence of Worcester (died AD 1118) 
in his work Chronicon ex Chronicis.} 

Silverman's online aurora catalogue\footnote{nssdcftp.gsfc.nasa.gov/miscellaneous/aurora}
does not list the event under those years AD 743 or 744
(see below), but mentions them in the commentary appendix 
for the mid 740s.
Fritz (1873) did not include this event explicitly in his extensive catalogue
(but see below and footnote 8),
even though of the previous publication by Jeremiah (1870) --
also not listed in later additions by Fritz.
Johnson (1880) -- listing {\em a red cross} from the ASC for AD 773 --
commented that there were three Schwabe cycles of $\sim 11.5$ yr each
between this event and a historic sun-spot observation from AD 807
(RFA) and concludes:
{\em It is probable an auroral light is referred to here}.
More recently, also Link (1962), Schlegel \& Schlegel (2011),
Usoskin et al. (2013), and Stephenson (2015) misinterpreted this sighting as aurora.

In the global catalogue of pre-telescopic astronomical observations 
by Hetherington (1996),
the event is also misinterpreted as aurora and listed three times 
namely for AD 773, 774, and 776:
Hetherington (1996) first cites Johnson (1880), 
who in turn cited the ASC ({\em red cross ... in sky}) for AD 773,
then Dall'Olmo (1979) for 
AD 774 (who did not list the {\em red cross}, 
except by quoting Link (1962) ({\em 774 (L)}), but Dall'Olmo (1979) did not give the text),
and finally, he cites Pang \& Bangert (1993) 
for quoting the ASC, but with a slightly different translation compared to AD 773
(now {\em sky} instead of {\em heaven}),
but the paper by Pang \& Bangert (1993) is unrelated, namely about a planetary conjunction 
several thousand years ago.

The phenomenon itself was transmitted in two lines, one by copying directly the ASC,
and one through secondary Latin compilations of historic events,
which mostly depended on the ASC (but not on version F), and which were appended somewhat.
This particular (Latin) version of the transmitted text does not include the wording
{\em a red cross/crucifix} (medieval English {\em Christesmel} and similar, 
Latin {\em crucis signum}), but instead has {\em rubea signa},
i.e. {\em red/reddish/brilliant signs} in plural:
Sch\"oning (1760) cites two sources in their original,
among them a chronicle from Joh. Broniton,
which are consistent in their content, 
but not in their dating (AD 776 and 773) --
and in this line of transmission, Sch\"oning (1760) can be seen as the first
who interpreted the event as aurora.
In his often used {\em Verzeichnis beobachteter Polarlichter},
Fritz (1873) gives just {\em Gross (Sch\"onning)} for AD 776, i.e. {\em Large},
but Sch\"oning (1760) was available to Fritz (1873) 
only as an excerpt in {\em Ch. U. D. Egger's Island},
as specified in Fritz (1873).\footnote{Fritz (1873) gives Sch\"onning 
or Schoenning instead of Sch\"oning, but means Gerhard Sch\"oning.
According to Fritz (1873), Sch\"oning compiled a catalogue of aurorae
from earlier reports with direct quotations, and his work would be much
more credible than others like Frobesius.}  
The citation as {\em Gross} ({\em Large}) is consistent
with the wording in Sch\"oning (1760):
{\em the northern light was again visible very large and terrible}
(translated by us from the original Danish).
The source for Sch\"oning (1760) was Matthaeus Westmonasteriensis:
{\em Anno gratiae 776 visa sunt in coelo signa rubea, 
post occasum Solis, \& horrenda.} 
({\em In the year of grace 776, red/reddish/brilliant signs were seen in the sky
after sunset, which were terrible/marvellous/astonishing.})
Sch\"oning (1760) translated and shortened {\em signa rubea} to {\em very large}
and {\em horrenda} to {\em terrible}.

The two lines of transmission merge in Link (1962),
who does not discuss those two lines, and the {\em serpents}
(actually adders) are connected here for the first time to
the interpretation as aurora.
In his aurora catalogue, Link (1962) gives the event for AD 773 (774)
following the uncorrected ASC, but quotes first 
the Latin secondary source on the same event from Roger de Hoveden (AD 1174-1201) 
from {\em Master of the Rolls} for AD 774 --
Roger did not mention the battle at Otford, which may have
resulted in incorrectly connecting {\em in coelo rubea signa}
with {\em serpentes ... in Sudsexe}.\footnote{Roger de Hoveden:
{\em Hoc autem anno visa fuerunt in coelo rubea signa post occasum solis horrenda:
serpentes visi sunt in Sudsexe cum magna admiratione} quoting from Link (1962),
who cited Roger de Hoveden from Masters of the Rolls 51/I, 23.} 
Link (1962) then gives as third report a quotation from
Matthaeus Westmonasterensis from {\em Flores Historiarum} --
meaning here Matthew Paris (died AD 1259), 
who had continued the chronicle of Roger of Wendover (died 1236);
these {\em Flowers of History} ({\em Flores}) also omitted the
battle at Otford after mentioning the sighting of the {\em signa rubea}
(see above), 
but listed those signs (probably by chance) in the correct year AD 776; 
it is interesting to note that this late but not independent source said:
{\em et serpentes visi sunt in Suthsexia, 
cum admiratione magna, ac si scaterent de terra},
i.e. {\em serpents/snakes coming out of the ground}.

As a summary, Link (1962) interprets this event as aurora
({\em C'est toujours la m\^eme aurore}), but he connected 
{\em rubea signa}, respectively a {\em red cross}, with {\em serpentes} 
({\em serpents}). Connecting these two parts is not justified, 
because they are obviously different events: In the older source, the ASC,
the battle at Otford is mentioned in between;
the entry reported three unusual facts for the year,
such a list is not atypical for a chronicle.\footnote{Gibbons \& Werner (2012) point
to the fact that the red colour (of the cross/crucifix) 
may be seen here as a {\em traditional motive for battle-related portents}.
Indeed, all three events may be seen as apocalyptic signs.
Usoskin et al. (2013) also point to the fact that 
{\em at this time the bible was a key reference in interpreting natural phenomena
explaining the cryptic reporting of aurorae}.
First, a {\em cross of light} itself seen in the sky or heaven was seen by Christian authors
at that time as positive portent, see e.g. the Chronicle of Zuqn\textit{\={\i}}n finished by a Christian
monk in AD 775 (Harrak 1999): A solar halo cross seen in the sky after an Earthquake
around AD 525/6 is reported as a sign of grace for the survivers,
interpreted as the help or presence of god --
while the aurorae in AD 772/3 were interpreted as negative portent 
(Neuh\"auser \& Neuh\"auser 2014, 2015a).
The red colour of the sign (cross/crucifix) in AD 776, as mentioned in the English versions of the ASC,
could be related to blood, battle, fire, etc. 
However, in that case, it is not possible 
to use an ideological criterion, i.e. the interpretation, to find out
whether the reported sighting ({\em red cross}) was seen positive (halo-like) or negative (aurora-like), 
because it is neither clear who has won the battle at Otford, 
nor who has reported the war and the sightings (Stenton 1970).}
Also, the wordings {\em in the South Saxon's land} (and {\em out of the ground})
do not speak for phenomena on the sky.
Link (1962) also notes for this event that it would give
the colour red ({\em r}) and also motion ({\em m}) --
but the motion is reported explicitly only for serpents/snakes on {\em the ground} 
({\em Flores Historiarum}). Link (1962) obviously thought that
serpents imply motion, see also Dall'Olmo (1980).

Link (1962) has searched for the original sources especially
for events from Fritz (1873). 
However, neither the {\em red signs} nor the {\em red cross} 
should be connected with {\em serpents} to one event.
Also the wording {\em rubea signa post occasum solis}
does not justify an interpretation as aurora,
because it is just a second line of transmission
from {\em red cross/crucifix} --
but such a sighting of a {\em cross} on the sky often
features in historic chronicles at that time,
but they were otherwise (correctly)
never interpreted as aurorae by Link (1962) nor others.

Silverman (1998, with online catalogue)
gives a listing of many sources, which may refer to
aurorae. He lists this {\em red cross} three times,
namely for AD 773 following Johnson (1880) and Britton (1937),
which both quote the ASC,
then for AD 774 citing Link (1962),
and also for AD 776 from Lowe (1870),
the latter gives {\em Brilliant} and {\em Aurora Borealis 
from the British Isles}, 
most certainly derived from the secondary Latin transmission
of {\em rubea signa}, which means something like a {\em reddish brilliant} display.
For other aurorae, Silverman does cite Newton (1972),
but for this events, he does not mention that Newton (1972)
noticed that the {\em red cross at 776} was 
a {\em refraction or reflection phenomen}[on].

In his large study on
{\em Medieval Chronicles and the rotation of the Earth},
i.e. mainly on historic reports of eclipses,
Newton (1972) does list quite a number of other celestial events 
from about the 8th to the 12th century -- so in appendix IX
{\em Meteorological Reports} sorted by regions:
Therein, he lists also atmospheric phenomena, identified by him as halo displays,
including the one dated (correctly) for AD 776.
For chronicles from Britain, he lists the following crosses
(brackets and numbering from us, see also next paragraph): \\
(1) {\em 776 red cross in sky after sunset,} \\
(2a) {\em 806 Jun 4} (full moon June 4 at UT 14:40h) {\em cross about the moon, 
gives a drawing, near dawn,} \\
(4?) {\em 1097 cross in the sky,} \\
(7) {\em 1156 Oct cross about the moon,} \\
(9) {\em 1191 sign of the cross with a crucified figure on it,} \\
(10) {\em 1208 May 3} (full moon May 2 at UT 5:12h) {\em crosses in the sky seen in Holland.} \\
Lunar phases here and below from F. Espenak, NASA\footnote{eclipse.gsfc.nasa.gov}.
For the sighting mentioned for AD 806 Jun 4, there is a drawing in the ASC,
see Fig. 2.

\begin{figure*}
\begin{center}
\includegraphics[angle=0,height=10cm,width=14cm]{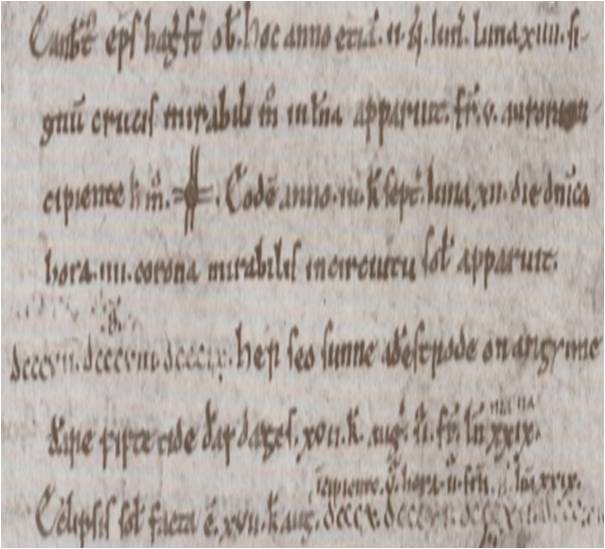}
\caption{{\bf AD 806.} A copy of a small part of the {\em Anglo-Saxon Chronicle} manuscript F,
here only the Latin part, with the
entry for AD 806, where the drawing of the {\em cross about the moon} is seen,
a lunar halo display.
The relevant text starts in the first line with the fifth word: {\em hoc anno etiam -ii- n[onas]
Iunii, luna -xiiii-, signum crucis mirabili m[odo] in luna apparuit f[e]r[ia] -v- aurora
incipiente, h[oc] m[odo] + Eodem anno -iii- K[alendas] Sept[embris], luna -xii-, die d[omi]nica hora -iiii-,
corona mirabilis in circuitu solis apparuit. 807 808 809 ...}. 
This (and the following as shown above) can be translated to (our addition in square brackets):
{\em Also in the same year, on 4 June [14th day of the moon], the sign of the holy cross appeared in the moon one
Wednesday at dawn; and again this year, on 30 August, a marvellous ring appeared around the sun.
807 808 809 In this year there was an eclipse of the sun at the beginning of the 5th hour of the day,
on 16 July, on Tuesday, the 29th day of the moon} (Latin and English text from Garmonsway 1953).
This means that there were no reports for the years 807 and 808 and that there was a solar
eclipse seen on AD 809 July 16, correctly dated (e.g. Newton 1972).
In the above scan, we can see the drawing of the observed phenomenon (written as + in the caption
text here), which is well consistent with an horizontal arc and a vertical pillar near or extending
from the moon.
Original at British Library, London, UK; this scan from the facsimile in Dumville \& Keynes (1995);
re-printed here with permission of the publisher Boydell \& Brewer Inc., UK.}
\end{center}
\end{figure*}

For mainland Europe, Newton (1972) lists more halo crosses,
four of which may correspond to the above cases,
so that there are at least ten independent reports: \\
(2a) {\em 806 Jun 4 cross about the moon near dawn,} \\
(2b) {\em 806 Jun 5? cross, cannot tell if sun or moon,} \\
(3) {\em 959 cross seen,} \\
(4) {\em 1096 Aug 7} (full moon Aug 6 at UT 19:50h) {\em a cross in the sky,} \\
(5) {\em 1152 Mar 22} (full moon Mar 21 at UT 13:00h) {\em circle around moon 
and rays like a cross reached from the moon to the circle,} \\
(6) {\em ca. 1155 3 moons, 3 suns, cross about the moon,} \\
(7?) {\em 1156 cross about the moon,} \\
(8) {\em 1157 3 moons with a cross in the center.} \\
In the lists in his appendix, there are many kinds of other
types of halos displays.
We can explain the relevant sighting following
Minnaert's text book {\em Light and Color in the Outdoors}
(Minnaert 1993), an earlier edition is already consulted by Newton (1972), 
almost always valid for both the Sun and the moon
(see Figs. 3-6): 
\begin{quotation}
The horizontal or parhelic circle: 
This is a circle running at the same height as the sun, 
parallel to the horizon. ... 
The fact of its being uncoloured shows clearly 
that it is caused by reflection, not refraction. ... 
Light pillars or sun pillars: 
A vertical pillar of light ... can be observed fairly 
often above the rising or setting sun ... 
This pillar of light is in itself uncoloured, 
but when the sun is low and has become 
yellow, orange, or red, the pillar naturally assumes 
the same tint. It is generally only 
about $5^{\circ}$ high, seldom more than $15^{\circ}$. ... 
Crosses: When a vertical pillar and a part of the horizontal 
circle occur at the same time, we see a cross in the sky. 
Needless to say, the superstitious have made the most of this!
\end{quotation}

\begin{figure*}
\begin{center}
%{\includegraphics[angle=0,width=17cm]{NationalGeographic_749109.jpg}}
{\includegraphics[angle=0,width=17cm]{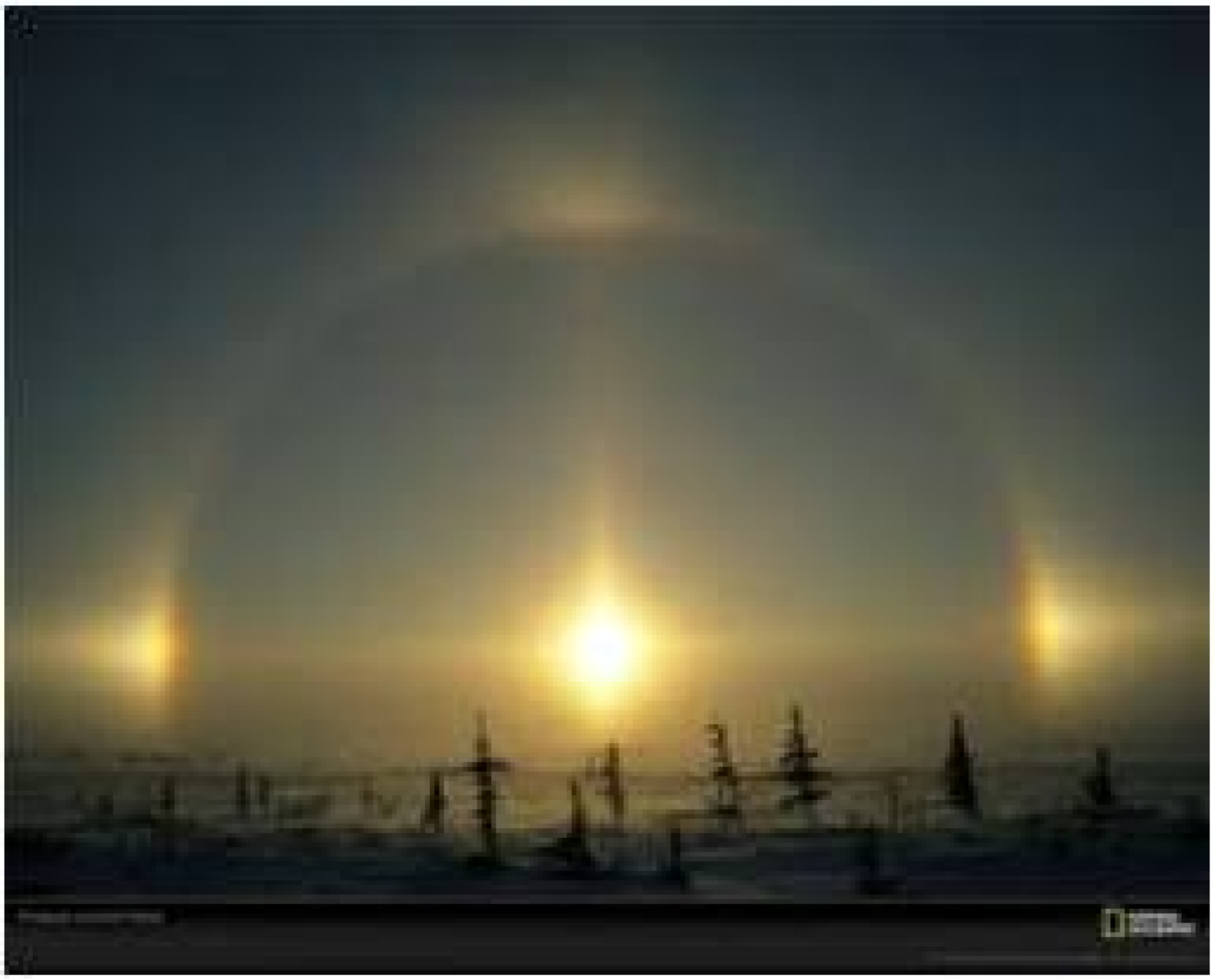}}
\caption{{\bf Solar halo cross.}
An image of the sun near horizon with mock suns (or sun dogs), the $22^{\circ}$ halo (circle),
horizontal arc, and vertical solar pillar, together forming what can be interpreted as cross
(reddish colour of pillar and horizontal arc due to reflection of a low altitude reddish sun).
Photograph taken by Norbert Rosing in Churchill, Manitoba, Canada; image copyright at National Geographics;
printed here with permission of National Geographics. Other good versions of parhelic halos
with and without an apparent cross can be seen at, e.g., www.atoptics.co.uk
or www.meteoros.de/halo.htm, the latter with German text.
Photographs with a halo cross around the moon can be seen, e.g., on
nyrola.jklsirius.fi/tmp/halot20041122/COMMON/100$\_$FUJI/TIDX0001.HTM,
www.atoptics.co.uk,
picasaweb.google.com/lh/photo/OX92bMlY3rbzpMR6aCMJYNMTjNZETYmyPJy0liipFm0?feat=directlink, 
or www.meteoros.de/halo.htm.}
\end{center}
\end{figure*}

\begin{figure*}
\begin{center}
{\includegraphics[angle=0,width=15cm]{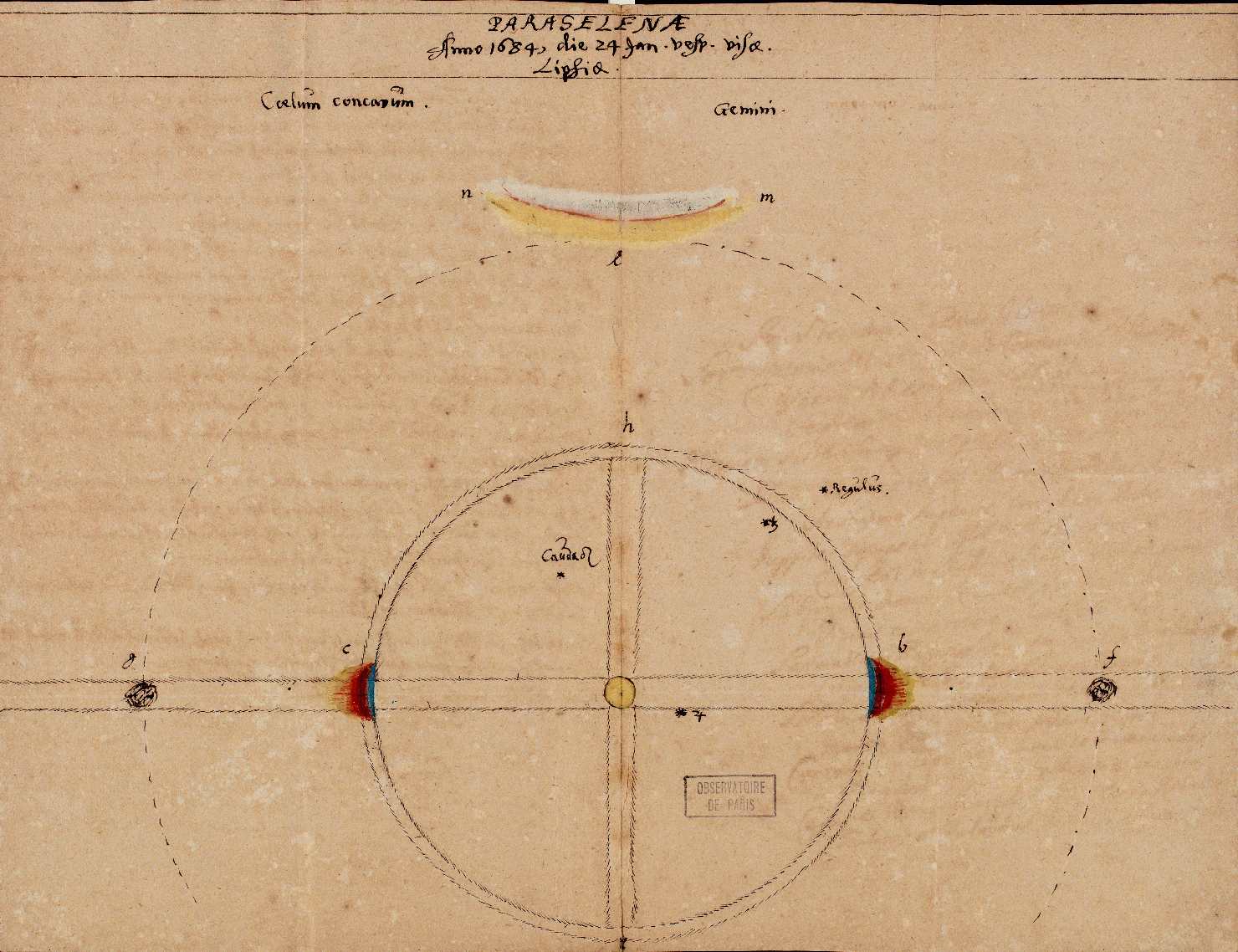}}
\caption{{\bf Paraselene.} 
A drawing by Gottfried Kirch (Leipzig, Germany, AD 1684) of the moon 
close to Jupiter (bottom center),
the $22^{\circ}$ (through b, h, c, and k) and $46^{\circ}$ (through f, l, and g) halos, 
a horizontal arc ($360^{\circ}$ through g, c, b, and f, Fig. 5), and a vertical pillar (from h to k) forming a cross.
Also seen are mock-moons of the $22^{\circ}$ (b, c with red colour) and of the $46^{\circ}$ (f, g) halos
and (probably) part of the circumzenithal arc (n, l, m) in or near the constellation of Gemini.
The caption text at the top says {\em PARASELENAE / Anno 1684, die 24. Jan. vesperi visae / Lipsiae}
(i.e. as seen in Leipzig in the evening of AD 1684 Jan 24, 
given in a Protestant area, which is Feb 3 on the Gregorian calendar)
and then {\em Coelum concavum}, i.e. the projection, 
the monn being in the south-east drawn in the center,
so that, analogously to Fig. 6, east is to the left and south to the right.
Gottfried Kirch has drawn this paraselene in a letter to Johannes Hevelius, Gdansk, Poland. 
The letter is reprinted as letter no. 256 in Herbst (2006) with the drawings
from the archive at Paris Observatory (BO Paris, C.1.16, no 29, folio 2299r-v and no 30, folio 2300). 
Kirch mentioned in the letter the following details:
the moon with cross was observed since 20:30h local time, mock-moons and 
the arc near the zenith had colours like a rainbow,  
the mock-moon g of the larger halo ring was elongated {\em like a comet} 
and observed only briefly, mock-moon f would have been seen briefly
in the same way by someone else;
at local midnight the halo display was gone (and it was very cold).
While the drawing is otherwise fully realistic (and possible),
the colours of the mock moons b and c are not drawn in the correct order,
but also not fully in rainbow order
(probably due to shortcomings in his memory late at night),
correct would be red at the (concave) inside for halos and outside (top) for rainbows. 
This image was received as digital copy from
Paris Observatory library and is shown here with kind permission.}
\end{center}
\end{figure*}

\begin{figure*}
\begin{center}
{\includegraphics[angle=0,width=12cm]{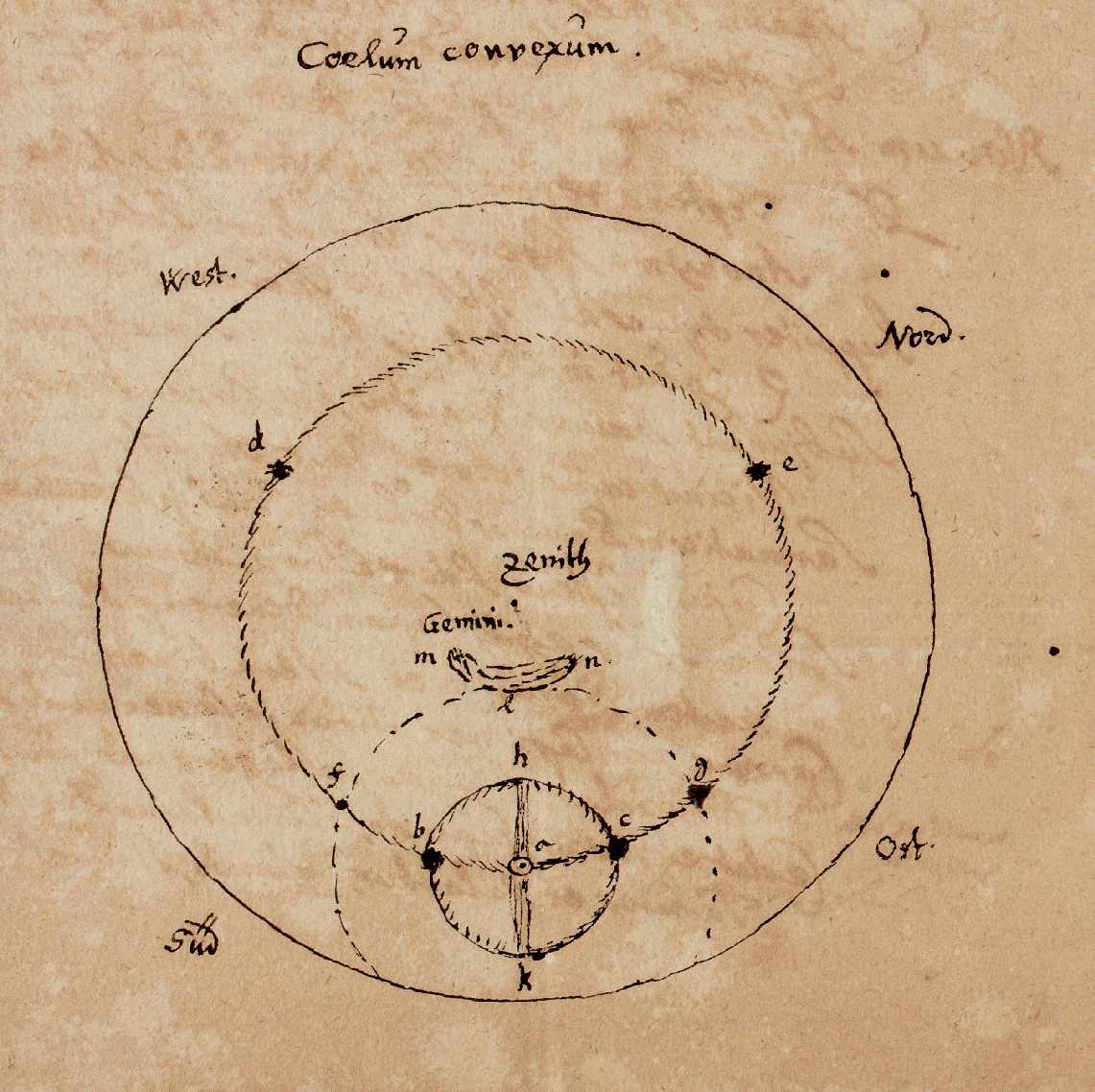}}
\caption{{\bf Paraselene.} Another drawing by Gottfried Kirch of the paraselene
seen on AD 1684 Jan 24 (Julian calendar) as shown in his letter to Hevelius from that month. 
We see the moon (bottom center, a), the horizontal arc being $360^{\circ}$
(through abcdefg), and also the features from Fig. 4 (four mock moons, two halo circles,
the vertical pillar forming a cross with the horizontal arc, and arc
below zenith and Gemini). The caption says {\em Coelum convexum}, i.e. the projection,
and then {\em West} (top left), {\em Nord} (top right for north),
{\em S\"ud} (bottom left for south), and {\em Ost} (bottom right for east) 
along the horizon drawn as circle. See also Fig. 4.
Kirch wrote that, at 23h local time, 
the horizontal arc reached all around the horizon ($360^{\circ}$).
Kirch: {\em In this halo circle, there were two mock moons d and e,
each of which one third of the circle away from the real moon,
rather large and clear, but not comparable with b and c ... this time,
together with the real moon and the six mock moons, there appeared seven moons.} 
The letter is reprinted as letter no. 256 in Herbst (2006) with the drawings, 
from the archive at Paris Observatory (BO Paris, C.1.16, no 29, folio 2299r-v and no 30, folio 2300). 
This image was received as digital copy from
Paris Observatory library and is shown here with kind permission.
}
\end{center}
\end{figure*}

\begin{figure*}
\begin{center}
{\includegraphics[angle=0,width=10cm]{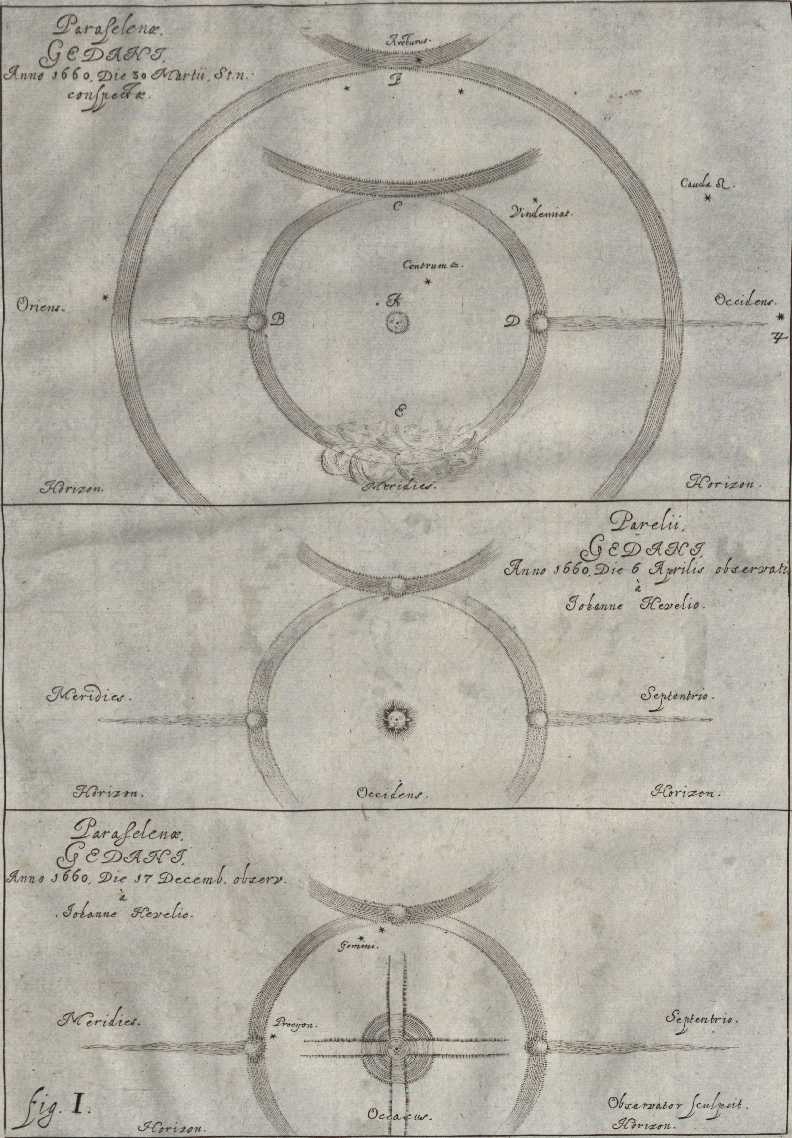}}
\caption{{\bf Paraselene.} Drawing by Hevelius (1611--1687, Gdanzk, Poland):
(a) Top. Paraselene with the moon (A) in the center and the paraselene mock moons of the
$22^{\circ}$ halo to the left (B) and right (D), 
both showing something like tails directed away from the moon,
(probably) the upper tangential arc of the $22^{\circ}$ halo (C),
some clouds (non atypical for halo displays) below the moon (E), 
the $46^{\circ}$ halo (touching the horizon), 
(probably) the cirumzenithal arc (F, top), as well as Jupiter (to the right, west) and a few stars.
His caption dates the sighting to the night of AD 1660 Mar 29/30 (Gregorian)
after midnight, i.e. some 74h after full moon.
(b) Middle. $22^{\circ}$ halo around the Sun (center) together with 
(probably) upper tangential arc (top) and two mock suns with tails
directed away from the Sun (towards the south, left, and the north, right),
seen on AD 1660 Apr 6 (Gregorian) in the evening in the west at low altitude by Hevelius,
who reported three extra suns ({\em tres solae}) and rainbow colours for the $22^{\circ}$ halo and the mock suns.
(c) Bottom. Paraselene with the moon in the center
and the horizontal arc and vertical pillar together forming a cross,
also seen are the mock moons of the $22^{\circ}$ halo
with something like tails directed away from the moon (towards the south, left, and north, right),
the inner $9^{\circ}$ halo,
and (probably) the upper tangential arc above the moon.
His caption dates the sighting to AD 1660 Dec 17 (Gregorian).
Hevelius described the sighting with {\em quattuor lunae duplici halone},
i.e. that he saw the moon plus three apparent moons: the real moon plus two mock moons 
plus the intersection of the $22^{\circ}$ halo with the upper tangential arc
counted as one of three apparent moons (together four moons, {\em quattuor lunae}) 
-- plus two halos {\em duplici halone}).
Hevelius added: {\em above the real moon, the was a very large, whitish gleaming or
silver cross, which is very rare. It was so bright and luminous, that it shone clearly
until sunrise.}  
Given that Hevelius saw the cross in the early morning of Dec 17,
it was just about half a day after full moon. 
The drawings are from Hevelius (1662), received in digital form from the
Thuringian University and State Library Jena, shown here with their kind permission.
}
\end{center}
\end{figure*}

In the list cited above from Newton (1972) there are
several reports about crosses explicitly mentioning the moon:\footnote{Newton (1972)
noticed also for the Sun as follows: 
{\em The configuration that seems to have been most striking to a medieval Christian 
is the one that is usually called ''the sign of the cross in the Sun'' 
in the medieval annals ... Many medieval sources contain drawings of this apparition.} }
Lunar crosses are possible only around full moon,
when the sky is usually too bright to detect aurorae --
this also speaks against the opportunity of detecting
both a cross around the moon and aurorae ({\em serpents/snakes}) in the same night.

It remains to be considered which kind of halo cross was meant in the ASC
reported for AD 776, a cross around the sun or the moon?
The medieval English texts have {\em \ae fter sunnan setlgonge}
(version A) or {\em setlgange} (B-E), the bi-lingual F also
in medieval Latin {\em post solis occubitum}.
Both the English {\em after/\ae fter} (Clark Hall 1960) and
the Latin {\em post} (Niermeyer 1976) had the meaning
of both our today´s {\em after} as well as {\em during/around}.
Hence, both is possible: a lunar cross {\em after} sunset 
(or, extremely seldom, {\em during} sunset) or
a solar cross {\em during} sunset. 
Furthermore, the wording {\em after sunset} tends to mean twilight, 
but does not mean in the dark night, so that an aurora is much less likely in any case.

All ASC manuscripts (except the Latin version of F) mention a red colour of the sign;
since versions A-E are older than F (F was finished until AD 1058),
the colour is quite trustworthy.
Because the cross was explicitly reported to be red,
it is more likely a cross during sunset (or moon rise).
Both, the parts of the horizontal arc as well as
the vertical pillar -- as being due to reflection -- have the same colour as the sun (or moon) 
depending on the height above horizon (e.g. Minnaert 1993):
the lower the more orange to red -- the higher. the more whitish. 
In principle, colours can be seen for lunar halo displays at night-time, 
like in case of lunar halo circles or mock moons.
The coloured light of halo features from the moon do not
activate the cone cells of human eyes so much,
because the moon is much fainter than the sun.
However, as seen in Fig. 4, red and yellow colours in mock moons (paraselene)
were clearly drawn and described by Kirch (in AD 1684).

A low-altitude moon appears more yellow-to-golden than red.
Also note that the medieval English word used in the ASC for this colour,
namely {\em read}, can mean {\em red of gold} (Clark Hall 1960).
Osterrieder (1995, page 179ff) also points to the fact that medieval manuscripts often
mean something like a {\em red-to-golden shine of light} when they use just {\em red}.
In any case: solar or lunar pillars (essential parts of the halo cross)
are much more probable during setting or rising of Sun or moon
than at higher altitudes -- so the reported information about the time
(after/during sunset) supports the interpretation of the sighting as halo cross.

In Newton's listing, there are more lunar crosses:
Among the ten events listed above, 
there are at least five which relate to the moon,
two of which have a dating (no. 2 and 5), namely within 1-2 days around full moon,
which shows how credible they are;
at two other reports (no. 4 and 10), which are dated, but where it is not mentioned whether
the cross was seen {\em about the moon} or the sun, the moon is also full;
for none of the ten sightings, the sun is given explicitly;
colour is mentioned only for the AD 776 event.

We conclude that the {\em red cross/crucifix} 
listed for AD 773, 774, or 776 was a solar or lunar
halo cross seen in AD 776. 

The sighting of the {\em red} halo cross in AD 776 was not the only one.
We would like to mention here another report about a {\em red cross}, 
which was seen not long after the one in AD 776:
According to a report from the monastery of San Juan de la 
Pe$\tilde{n}$a in Arag\'on, Spain, a red cross ({\em una cruz roja}) of light 
above a tree was seen by King Garci Jim\'enez 
at the beginning of the 9th century, namely during a battle against the Moors
at Ainsa in Aragon, Spain;
Garci Jim\'enez won the battle and then he founded the order of the knights 
of St. John in the monastery San Juan, who show the red cross on their coat of arms
(Osterrieder 1995 citing Martinez 1620). 

Our analysis of the sources and their transmission as well as their critical dating and context
(pointing to a halo display in AD 776 as in Newton 1972) also shows that
''{\em a quick Google search}'' (J. Allen as quoted by 
R.A. Lovett\footnote{www.nature.com/news/ancient-text-gives-clue-to-mysterious-radiation-spike-1.10898
dated 27 June 2012})
as done by Allen (2012) and as published in Nature can be insufficient.

Most recently, also Stephenson (2015) -- investigating historical records 
around AD 774/5 -- mentioned the {\em red cross} and concluded that it is
{\em plausible} ... {\em a display of the aurora}, because it was {\em red} and {\em after sunset};
he dated it to AD 773 or 774 or 776, partly based on late secondary compilations,
but he did not discuss the two-year shift in the ASC nor the possible meanings
of the medieval English {\em \ae fter} around that time, 
nor did he consider a halo display 
(even though he cited Newton (1972), who classified it as a halo display in AD 776). 

\section{''Inflamed shields'' in the sky}

Usoskin et al. (2013) listed {\em ''inflamed shields'' in the sky} for AD 776
and an {\em apparition interpreted by Christians as riders on white horses} for AD 773, both allegedly aurorae,
both from the Royal Frankish Annals (RFA), both about victories of the Frankish army against the Saxons.
Let us first discuss the former event (and in the next section the latter).

In the English translation of the RFA by Scholz \& Rogers (1970), 
we can read about the first event and its circumstances as follows: 
\begin{quotation}
AD 776 ... Then a messenger came with the news that the Saxons had rebelled, ...
With [castle] Eresburg thus deserted by the Franks, the Saxons demolished 
the buildings and walls. Pathing on from Eresburg they wished to do the same thing to
the castle of Syburg, but made no headway since the Franks with the help of God put up a manly resistance.
...
But God's power, as is only just, overcame theirs. One day, while they prepared for battle
against the Christians in the castle, God's glory was made manifest over the castle church
in the sight of a great number outside as well as inside, many of whom are still with us.
They reportedly saw the likeness of two shields red with flame wheeling over the church.
When the heathens outside saw this miracle, they were at once thrown into confusion and started
fleeing to their camp in terror ... 
But the more the Saxons were stricken
by fear, the more the Christians were comforted and praised the almighty God who deigned to
reveal his power over his servants.
\end{quotation}
In the original Latin, the text reads as follows (from Rau 1955): 
\begin{quotation}
AD 776 ... Tunc nuncius veniens, qui dixit Saxones rebellatos, ...; 
sic Eresburgum a Francis derelictum, muros et opera destruxerunt.   
Inde pergentes voluerunt de Sigiburgi similiter facere, auxilianter Domino 
Francis eis viriliter repugnantibus nihil praevaluerunt. 
...
Sed Dei virtus, sicut iustum est, superavit illorum virtutem, et quadam die, cum bellum praeparassent adversus
christianos, qui in ipso castro residebant, apparuit manifeste gloria Dei supra domum ecclesiae, quae est infra ipsum
castrum, videntibus multis tam aforis quam etiam et deintus, ex quibus multi manent usque adhuc; 
et dicunt vidisse instar duorum scutorum colore rubeo flammantes et agitantes supra ipsam ecclesiam. 
Et cum hoc signum vidissent pagani, qui aforis erant,
statim confusi sunt et magno timore perterriti coeperunt fugere ad castra ...
attamen quantum illi plus pavore perterriti fuerunt, tanto magis christiani confortati omnipotentem
Deum laudaverunt qui dignatus est suam manifestare potentiam super servos suos. 
\end{quotation}

Part of the story above (starting after {\em praevaluerunt} until the end of our quotation)
is missing in three manuscripts,
and it is inserted at various different locations in other manuscripts (Rau 1955).
This story is a comment found at the side of the manuscript.\footnote{For details,
see Neuh\"auser \& Neuh\"auser (2014), footnote 33.}
However, this does not speak against its reliability.
Rau (1955) and Scholz \& Rogers (1070) inserted this comment at the same location,
which does fit well given the context.

What Usoskin et al. (2013) describe as {\em ''inflamed shields'' in the sky}
are given as {\em the likeness of two shields red with flame wheeling over the church}
in the translation by Scholz \& Rogers (1970) cited by Usoskin et al. (2013). 

Apparently, the event took place {\em during the day} (Latin {\em quadam die}), 
as already noticed by Gibbons \& Werner (2012).
Pilgram (1788), however, assumed that the event should have taken place during the night,
so that he could classify it as aurora -- dated by him to AD 775, even though his
source, Sigebert of Gembloux, dated it to AD 776.
Neither Link (1962) nor Silverman (online catalog) nor Usoskin et al. (2013),
who both followed Link (1962), considered the day-time.

Link (1962) gave the {\em Annales Laurissenses} as source, i.e. the RFA, and claimed
without giving additional evidence that {\em cette relation se trouve \'egalement dans le Chronicon Sigeberti G.}
and Lycosthenes. The problem, however, is that those two reports are not identical:
While Sigebert (died AD 1112) and then -- depending on him -- Lycosthenes (died AD 1561) connected
the {\em two shields} with the castle called Eresburg (today Marsberg-Eresperg in Germany),
a misidentification going back to Chronicon Regino (about AD 906),
the RFA connected them with castle Syburg near today's Dortmund in Germany.
In RFA, it is mentioned that castle Eresburg was recaptured by the pagan Saxons;
given that the RFA was written under and for Charlemagne, 
any reports being negative for the Franks indicate high credibility. 
Then, it is reported that castle Syburg could be
hold by the Franks {\em with the help of God},
and -- as given in the comment -- 
due to {\em shields ... over the church}, a presumable intervention 
by god, which in turn, in their belief, facilitated their victory.

For completness, Dall'Olmo (1978) interpreted this event 
(Lat. {\em instar duorum scutorum flammantes} presumable from
Annales Bertiani, an extension of the RFA which started later, 
hence he probably meant the RFA) as two bright meteors or as aurora.
Hetherington (1996) also gave the Annales Bertiani quoting Dall'Olmo (1978) for this event.
Also, Stephenson (2015) follows Dall'Olmo (1978) in the interpretation of this event.

The report about {\em the likeness of two shields red with flame wheeling over the church}
(Latin: {\em instar duorum scutorum colore rubeo flammantes et agitantes supra ipsam ecclesiam})
used the word {\em instar} for {\em something like} or {\em likeness}, 
i.e. not really two red shields, but something {\em like} two red shields,
which were located above the rooftop of the church. A comparison with two shields may indicate a roundish form,
definitely {\em two} similar phenomena {\em over the church} -- indicating similar direction and
(or at least being consistent with) similar or same height above the horizon -- and then 
{\em two shields red with flame wheeling}, i.e. with both red colour and some flame-like behaviour.
Both the Franks as well as the Saxons saw the event.

Mock suns fullfill all elements of the description, Minnaert (1993): 
\begin{quotation}
{\em The parhelia or mock suns (sun dogs) of the small halo: 
These mock suns are two concentrations of light on the small halo ($22^{\circ}$ halo) at the same altitude as the sun. 
It often happens that only one of the two can be seen properly and sometimes the small halo is absent, 
whereas the parhelia (mock suns) are clearly visible. 
The intensity of these mock suns is usually very great; 
they are distinctly red on the inside, then yellow, before changing into a bluish white.}
\end{quotation}
Such red mock suns are seen in Fig. 3 (red mock moons in Fig. 4). 
Especially intensely shining red parhelia resemble 
something like {\em flames} (often described like that in historic literature) 
regarding colour and the impression of motion.

Schreiber (1984) also interpreted this sighting as some kind of a solar light effect.
What is reported as {\em the likeness of two shields red with flame wheeling over the church} 
clearly were two mock suns. Given that nothing was reported about an ongoing fight, 
it appears more likely that the event took place during the morning.
The Saxons were in full armament outside of the camp near the Frankish castle ready to attack.
Then, they saw the Franks celebrating the {\em two shields} as intervention 
by (their) god -- and fled panic-stricken.\footnote{Excavations 
on the Syburg plateau indicate that the church St. Peter (above which the sighting was reported) was
located in the southern part of the castle plateau (see www.syburg.de/sy-plan1.htm).}

Link (1962) also remarked that there was motion ({\em mouvement observ\'e}) seen in this event
(in addition to red colour).
It may well be that either {\em flammantes} and/or {\em agitantes} may have led him to suppose motion.
The latter -- together with the location, here a church -- can 
also mean {\em abidance} or {\em linger around} (translated from German following Petschenig 1971),
which would fit well with mock suns. Otherwise, {\em agitantes} can be interpreted as duplication of  {\em flammantes},
because something described as {\em flaming} also appears as moving or wheeling
(however, staying {\em over the church}).
Apparent motion is not atypical for mock suns, because they change their appearance with the motion of
the cirrus clouds, which can move and/or change fast. 
Also, we should not forget the situation -- people on the move
immediately before an attack and/or a counter-strike -- or fleeing.

This story is an example for the predominantly positive interpretation
of halo phenomena by Christians, namely as positive portents.

\section{''Riders on white horses''}

Usoskin et al. (2013) {\em in a new survey of occidental chronicles ... identified
probable aurorae ... in AD 773
an apparition interpreted by Christians as riders on white horses (Germany)}
-- an event which was never before suggested as aurora.
Usoskin et al. (2013) cited Scholz \& Rogers (1970),
who translated the apparition as {\em two young men on white horses}.

The event is narrated in the RFA as follows (translation by Scholz \& Rogers 1970): 
\begin{quotation}
AD 773 ... The Lord Charles [Charlemagne] celebrated Christmas [Dec 773]
there [Pavia, Italy] in his camp, and he celebrated Easter [774 Apr 3] in Rome.
While he went to Rome during this year to defend God's Holy Roman Church
at the invitation of the supreme pontiff,
the borderline against the Saxons was exposed and not secured by any treaty. 
The Saxons, however, fell upon the neighbouring Frankish lands with a large army
and advanced as far as the castle of B\"uraburg. The inhabitants of the borderland were
terrified when they saw this and retreated into the castle. When the Saxons in their
savagery began to burn the houses outside, they came upon a church at Fritzlar
which Boniface of saintly memory, the most recent matyr, had consecrated and
which he had said prophetically would never be burnt by fire. The Saxons began to attack
this church with great determination, trying one way or another to burn it.
While this was going on, there appeared to some Christians in the castle and also to
some heathens in the army two young men on white horses who protected the church from
fire. Because of them the pagans could not set the church on fire or damage it, either inside
or outside. Terror-stricken by the intervention of divine might they turned to flight ...
And the date changed to AD 774 ...
\end{quotation}

In the original Latin, the text is as follows (Rau 1955):
\begin{quotation}
AD 773 ... Ibique domnus Carolus in sua castro natalem Domini celebravit
et pascha in Roma. Et dum propter defensionem sanctae Dei Romanae ecclesiae
eodem anno invitante summo pontifice perrexisset,  
dimissa marca contra Saxones nulla omnino foederatione suscepta.
Ipsi vero Saxones exierunt cum magno exercitu super confinia Francorum, 
pervenerunt usque ad castrum, quod nominatur Buriaburg;
attamen ipsi confiniales de hac causa solliciti, cumque hoc cernerent, castello sunt ingressi. Dum igitur
ipsa Saxonum gens coepisset serviens domos forinsecus incendia cremare, venerunt ad / quandam basilicam in loco,
qui dicitur Friedislar [Fritzlar], quam sanctae memoriae Bonefacius novissimus martyr consecravit atque
per spiritum prophetiae praedixit, quod nunquam incendio cremaretur. Coeperunt autem idem praefati Saxones
cum nimia intentione adversus eandem certare basilicam, quemadmodum eam per quodlibet ingenium ingi
cremare potuissent. Dum haec igitur agerentur, apparuerunt quibusdam christianis, qui erant in castello,
similiter et quibusdam paganis, qui in ipso aderant exercitu, duo iuvenes in albis, qui ipsam basilicam
ab igne protegebant; et propterea ibidem non potuerunt neque interius neque exterius ignem accendere
nec aliquod dampnum eidem inferre basilicae, sed nutu divinae maiestatis pavore perterriti in fugam conversi sunt,
... Et inmutavit se numerus annorum in 774.
\end{quotation}

However, the story above (except the first and last sentence) about the church in Fritzlar is missing in manuscript A, 
and it is inserted at various different locations in other manuscripts (Rau 1955).
This does not mean that it is not reliable.\footnote{See Sect. 3 and footnote 14.}

In the main text, it is stated that Charlemagne celebrated Christmas in northern Italy 
and Easter in Rome (April 774).
The extra text narrates what happened during that time in Germany. 
Hence, the events probably took place in winter and/or spring.
The turn of the calendar year in Charlemagne's empire at that time (until AD 800) was at Easter.
So, it could be that the events took place still in their year 773;
but according to our calendar system and our definition of the turn of the year,
it most likely happened in AD 774.

The Latin wording {\em duo iuvenes in albis} can mean either {\em two young men on white horses} 
or {\em two young men in white} (clothes). 

For better understanding the kind of sighting which is reported in RFA,
we refer to another source with information about these events:
the {\em Vita} of Wigbert (died AD 746 or 747), holy for the Christian church, 
who was an Anglo-Saxon monk working in central Europe, like Boniface (died AD 754);
the latter installed the former as prior of the monastery of Fritzlar, the town mentioned in the report.
In that {\em Vita}, in the context of an apparition, it is explicitly mentioned 
that the Saxon army was in its camp during the night 
(where they should be normally at night): 
\begin{quotation}
the Saxons wanted to fight ... but in the night prior to the day of the execution of their cruel aim,
they suddenly got very much frightened, they saw that the above mentioned church [of Fritzlar] was attached to
something like a bright flow from heaven and with a noble figure going back and forth on its roof ridge,
having streak and contour of a human body,
but being much larger than humans in impressiveness and dignity and wearing white garment.
Full of horror about this sign of god and this large appearance ... and they flew.
\end{quotation}
Latin and Germans texts in Fleck (2010) translated to English by us; 
the original Latin runs as follows:
\begin{quotation}
... Saxones ... ipsa nocte, cui tam crudelis negocii dies illucere sperabatur,
incredibili repente perculsi formidine fusa caelitus clarissima luce ambiri conspicantur
aecclesiam, cuius memoria supra iam facta est, et per illius culmen insignem quendam
liniamentis quidem humani corporis circumscriptum, at valentiam dignitatemque mortalium longe
prestantem, albis etiam vestibus amictum hac et illac deambulare. Divino itaque
nutu et huius visionis exterriti maiestate ... ad fugam se denuo contulere.
\end{quotation}

Fleck (2010) supposed that the author of the Vita of Wigbert, written AD 836 by Lupus Servatus, 
knew and utilized the RFA.
That Lupus Servatus mentioned only one {\em noble figure} (Wigbert) 
instead of two ({\em two young men on white horses} in the RFA),
could be due to the fact that he wanted to focus on Wigbert (Fleck 2010). 
While some wordings in the RFA and the Vita of Wigbert are similar, 
there are also clear differences in content:
in the Vita of Wigbert, the sighting is reported explicitly for night-time, 
the attack by the Saxons does not happen, 
and they try to set the church on fire later in different reports (no. 19 \& 21 in Fleck 2010),
while similar to the RFA (with Boniface instead of Wigbert), 
however, without narrating a special halo miracle; 
in those two reports in the Vita (no. 19 \& 21), the time of the day is not mentioned,
which may speak for day-time.

The appearance of {\em two young men on white horses} or {\em in white} clothing 
from the RFA is not directly reported in such a way in the Vita of Wigbert (and viceversa). 
To assume (Fleck 2010) that one of the two men (Boniface) was just omitted in order to focus on Wigbert
is not convincing, because Boniface was very much adored in that area at that time; 
it would not at all be negative for Wigbert to be mentioned together with Boniface. 
It is more likely that two different halo displays are reported,
once a lunar halo display at night (Vita of Wigbert: {\em something like a bright flow from heaven}) 
and once a solar feature in the RFA.

Maybe, the two reported halo miracles, a very serious event for Christians,
happened within a few days;
there are many examples, where lunar and solar halos occurred within such a short period,
given the prevailing meteorological conditions. 

What was reported as {\em something like a bright (white) flow from heaven} (at night) etc. 
could well be a vertical lunar pillar  
and maybe additionally a horizontal arc and mock moons forming something
like a figure, maybe with open arms and hands (and, hence, similar to a cross)
around full moon. 
Our interpretation of the sighting as halo effect is supported by the use
of the term {\em white figure of a man (homo candens) on the sky} in medieval Latin texts (Dall'Olmo 1980),
which was several times misinterpreted as aurora sighting --
however, as far as we know, this report with Wigbert was never suggested to be an aurora;
there are other examples given in Neuh\"auser \& Neuh\"auser (2014),
see also footnote 12 in Sect. 2.

The report of {\em two young men on white horses} (or {\em in white} clothing) could well be two mock suns,
possibly still elongated by parts of the horizontal arc and/or the $22^{\circ}$ halo circle, 
maybe also with a Lowitz arc; see also Sect. 5 for the motive of {\em young men on white horses}.
The phenomenon being reported here as {\em white} is fully consistent with parhelia,
which often appear only white or bright, even though mock suns refract sun light.

We also refer to a drawing of a solar halo display as observed by Hevelius (Fig. 6), 
where he explicitly describes {\em rather long and white glowing tails}. 
Minnaert (1993) adds about mock suns as follows: 
\begin{quotation}
On close observation, you will 
find that in reality the parhelia stand a little way 
outside the small halo, the more so as the altitude of the sun is greater; 
when the sun is very high, the difference may even amount to several degrees.
\end{quotation}

Our halo interpretation is not inconsistent with the report that both
{\em some Christians in the castle} (B\"uraberg) {\em and also to some heathens in the army}
(Saxons near Fritzlar) saw the phenomenon (not necessary by the very same crystals), 
even though they may have been separated by some distance and the river Eder:
halo sightings often happen above a river due to water vapour turning to ice crystals.

The castle B\"uraburg/Buriaburg, now named Ungedanken, 
today WSW part of the town Fritzlar in Germany: 
if the sighting and the report in the RFA were made by people in the castle B\"uraburg/Buriaburg looking
towards the church in Fritzlar (located across the river Eder), then they were observing 
towards the east.

\section{Ideological background of the motive ''two young men on white horses''}

Our interpretation of the sighting (Sect. 4: {\em duo juvenis in albis}) 
as mock suns is supported by the frequent use
of the antique {\em dioscuri motive} adapted by Christians in medieval times,
for which we explain the ideological background here.
We also give a few similar examples.

Theodoret, church historian (died about AD 457) 
reported about a similar simultaneous early morning sighting of mock-suns
by both the Christian-Roman Emperor Theodosius I. (AD 347-395) and a soldier, 
which reportedly motivated them to start the decicive attack on AD 394 Sep 6 at 
the battle near the river Frigidus in western Slovenia, often regarded as the final victory of Christianity
(mock moons are less likely because new moon was a few days later on AD 394 Sep 11 at 22h).
The strategic situation was not advantageous
for the Emperor's army the day before, so that the chief advisers considered to retreat; Theodosius 
\begin{quotation}
remained the whole night in prayer to the Lord of the world, 
namely in a small prayer hat, which he had found on a hill at his camp.
Around the time of cock row, however, he fall asleep\footnote{The Greek word 
here is {\em h\'ypnos} which also means {\em doze};
Theodoret reported soon later that the Emperor continued to pray even more intensively, 
so that the Emperor was probably not fully asleep.}. 
While he was lying on the ground, it appeared to him as if he would see two young men in white clothes
and sitting on white horses, who were supporting him to be of good courage, to banish fear from his heart, 
and amour and deploy the army at dawn.
They said to have been sent to help and to lead the attack.
One said he would be John, the evangelist, and the other one said to be Phillip, the apostle.
After this vision\footnote{The meaning of the Greek {\em \'o}$\psi \iota \varsigma$ (\'opsis), 
i.e. the Latin visio, gives preference to the a real sighting (Weber 2000).} 
the Emperor did not stop suppliant praying ...
The same vision was seen by a normal plain soldier, 
who told about it to a captain,
who informed the colonel, who in turn informed the general, and the latter informed the Emperor,
thinking that he would tell him important news.
\end{quotation}
(translated by us from the German translation from the original Greek in Seider 1926).

Afterwards, it is mentioned that the Christians won the battle 
partly due to strong winds blazing against the enemy,
which was seen as confirmation of the heavenly miracle;
meteorologically, there could be a connection between the storm  
and a cold front, which facilitates the formation of ice crystals for the halo effect; 
the still warm river Frigidus provided high humidity.
Both in this story from AD 394 as well as in the report discussed 
in Sect. 4 from AD 774, we have a victory of the Christians
supported by their interpretation of (the sighting of) 
{\em two young men in white clothes} and/or {\em on white horses}.
Weber (2000) remarked about the AD 394 story: 
{\em This description, in particular the two horses, shows that this is the first adaptation
of the ... dioscuri motive as an 'interpretatio Christiana'.}

Where does the {\em dioscuri motive} come from?

An earlier Roman story about the apparition of dioscuri, cited very often throughout antiquity, 
reports the victory of the Roman dictator A. Postumius Albus 
against the Latins at lake Regullus in Italy in BC 499 or 496:
\begin{quotation}
Among the riders, there appeared two young men on white horses fighting with extraordinary courage.
The dictator ordered to search for them, in order to honour them with gifts,
which were sufficiently worthy for them, but they were not found:
Postumius thought they were Castor and Pollux and dedicated a temple for the two.
\end{quotation}
(translated by us from the German in Weber 2000) -- again the sighting of two mock suns 
(since the fightings happened during the bright day, they were mock {\em suns} 
rather than mock {\em moons} or even real stars).\footnote{Also in other instances, 
mock suns were referred to as {\em stars}, e.g. Cassio Dio (AD 150-235) wrote 
for the year AD 193 about another sighting of mock suns next to the Sun: 
{\em This was what went on in Rome. Now I shall speak about what happened outside and the various revolutions.
There were three men at this time who were commanding each three legions of citizens and many foreigners besides,
and they all asserted their claims -- Sever\-us, Niger, and Albinus. The last-named governed Britain,
Severus Pannonia, and Niger Syria. These were the three persons darkly indicated by the three stars that suddenly
came to view surrounding the sun, when Julianus in our presence was 
offering the Sacrifices of Entrance in front of the senate-house.
These heavenly bodies were so very brilliant that the soldiers kept 
continually looking at them and pointing them out to one another,
declaring moreover that some dreadful fate would befall the usurper. 
As for us, however much we hoped and prayed that it might so prove,
yet the fear of the moment would not permit us to gaze at them, 
save by occasional glances. Such are the facts that I know about the matter.} 
(English text from 2004 edition of Cassio Dio on 
www.gutenberg.org/files/10890/10890-h/10890-h.htm$\#$b74).
While there exist only two $22^{\circ}$-mock suns, the {\em third star}
can be interpreted as the intersection of the $22^{\circ}$ halo with
the upper tangential arc, see Fig. 6b and c.}

Those {\em dioscuri}, which literally means {\em sons of Zeus} or 
{\em son of the god of heaven}, 
were Castor and Pollux for the pagan Romans,
they were John and Phillip for the Christian-Roman Emporer Theodosius, 
and they were obviously identified with Boniface and Wigbert by the Christian Franks
in the late 8th century. In antiquity, they are often drawn or sculptured as riding on horses;
in particular, they were almost always displayed together with horses. 
It is quite plausible that what Hevelius (1662, Fig. 6)
described as {\em rather long and white glowing tails} (for AD 1660 Apr 6),
the {\em tails} of the mock-suns, directed away from the Sun,
was interpreted as {\em tails} of horses, on which the two dioscuri were fighting.

\section{Summary}

The three sightings in the mid AD 770s interpreted as aurorae by Usoskin et al. (2013)
-- supporting their super-flare hypothesis -- 
(two of them also as aurorae in Link 1962 and others) were shown here to be halo displays.
Many similar sightings were reported in medieval times by Christians (crosses, circles, mock-suns, etc.).
They usually interpreted halos or similar displays 
as positive portents, as appearance or intervention by god (epiphany),
motivated by the Christian Bible.
On the contrary, auroral features were usually interpreted by Christian observers as negative portents.
See Neuh\"auser \& Neuh\"auser (2014) for more details and Neuh\"auser \& Neuh\"auser (2015a)
for aurorae examples and their Christian interpretation for AD 772 and 773.

The two sightings from the RFA 
cited by Usoskin et al. (2013) as {\em inflamed shields} and {\em riders on white horses}
were mock suns. The latter was narrated as a frequent Christian adaption of the well-known 
Roman dioscuri motive. The former ({\em two shields red with flame}) were probably interpreted by the
Christians as the shields of two heavenly fighters. 
The sighting of the {\em red cross} or {\em crucifix} in the ASC (AD 776), 
which was a halo display with horizontal arc and vertical pillar,
is not clearly given together with an interpretative context as the two others.
Other, very similar sightings allow an interpretation:
Christians identified it (the cross) as the proclaimed (Mt 24 in the Bible) sign of the so-called
{\em Son of Man}, thought to come at the end of all times (see Neuh\"auser \& Neuh\"auser 2014).
The entry of the ASC (southern events) offer an apocalyptic selection: cross -- battle -- serpents.

Note that the {\em inflamed shields}, the {\em riders on white horses}, the Wigbert story, 
as well as the sighting of Theodosius are sightings over Christian churches or chapels --
all sightings were interpreted as intervention by God and happened at
locations where the presence of God was assumed.

An additional alleged aurora reported by the Chinese -- presumably 
for AD 775 (Usoskin et al. 2013, Zhou et al. 2014) -- was actually in AD 776 Jan,
and possibly also some halo effect ({\em above the} (almost full) {\em moon}) instead of an aurora
(see Neuh\"auser \& Neuh\"auser 2015a and Chapman et al. 2015).
Given that the sightings were misinterpreted as aurorae in Usoskin et al. (2013) 
and given that two of the European events
were in AD 776, i.e. after the $^{14}$C increase from AD 774 to 775, 
they are anyway not related to the
cause of the $^{14}$C increase -- and in particular they do not support a super-flare hypothesis.

\acknowledgements
We obtained the data on moon phases from Fred Espenak, NASA/GSFC, available on eclipse.gsfc.nasa.gov.
We consulted the Silverman aurora catalog on nssdcftp.gsfc.nasa.gov/miscellaneous/aurora.
We would like to thank Thomas Posch (U Vienna) for accessing the book of Pilgram, which appeared
in 1788 in Vienna, which we could consult in their university observatory library.
We also would like to thank Regina von Berlepsch (AIP Potsdam) for providing us with a copy of the paper by Sch\"oning (1760),
whose relevant parts were then translated from old Danish by Sven Buder (U Jena).
We also acknowledge D. Luge and S. Daub (U Jena) for help with translation of medieval Latin texts.
We thank T. Honegger (U Jena) for advice on the meaning of {\em \ae fter} in medieval English.
We are thankful to Wendy Ellis from the publisher Boydell \& Brewer Inc., UK,
for their kind permission to scan and show copies of the Anglo-Saxon Chronicle facsimile 
published in Dumville \& Keynes (1995), whose original lies in the British Library.
We got the medieval English texts from of the Anglo-Saxon Chronicle (version A-E) from asc.jebbo.co.uk/b/b-L.html.
We thank Susan G. Henry from National Geographic Stock, Washington, D.C., USA, for permission
to present Fig. 3, and we also acknowledge Norbert Rosing, the photographer of the parhelic circle with cross.
We also thank Emilie Kaftan,
Biblioth\`eque de l'Observatoire de Paris, France,
for the digital copy of the letter from Kirch to
Hevelius with colour drawings.
The drawing from Hevelius was received in digital form from the 
Th\"uringer Universit\"ats- und Landesbibliothek Jena,
the Latin text from Hevelius was translated by D. Luge (U Jena).
Finally, we are grateful to our referee, Prof. J. Feitzinger, 
for his encouraging remarks.

\end{document}